# Characterization of a Canonical Helicopter Hub Wake

Christopher E. Petrin[*], Balaji Jayaraman[#] & Brian R. Elbing[†,1]

School of Mechanical & Aerospace Engineering
Oklahoma State University, Stillwater, OK 74078, USA

[*]ORCID: 0000-0002-4582-2679

[#]ORCID: 0000-0001-5825-6453

[†]ORCID: 0000-0002-0818-7768

## Abstract

The current study investigates the long-age wake behind rotating helicopter hub models composed of geometrically simple, canonical bluff body shapes. The models consisted of a 4-arm rotor mounted on a shaft above a 2-arm (scissor) rotor with all the rotor arms having a rectangular cross-section. The relative phase between the 2- and 4-arm rotors was either 0° (in-phase) or 45° (out-of-phase). The rotors were oriented at zero angle-of-attack and rotated at 30 Hz. Their wakes were measured with particle-image-velocimetry within a water tunnel at a hub diameter based Reynolds number of 820,000 and an advance ratio of 0.2. Mean profiles, fluctuating profiles and spectral analysis using time-series analysis as well as dynamic mode decomposition were used to characterize the wake and identify coherent structures associated with specific frequency content. The canonical geometry produced coherent structures that were consistent with previous results using more complex geometries. It was shown that the dominant structures (2 and 4 times per hub revolution) decay slowly and were not sensitive to the relative phase between the rotors. Conversely, the next strongest structure (6 times per hub revolution) was sensitive to the relative

---

[1)] Author to whom correspondence should be addressed. Tel: 405-744-5897; e-mail: elbing@okstate.edu

phase with almost no coherence observed for the in-phase model. This is strong evidence that the 6 per revolution content is a nonlinear interaction between the 2 and 4 revolution structures. This study demonstrates that the far wake region is dominated by the main rotor arms wake, the scissor rotor wake and interactions between these two features.

# 1  Introduction

The flow associated with the rotor hub of a rotorcraft (e.g. helicopter) has been a recognized problem for decades (Keys & Wiesner, 1975; Reich et al., 2016) due to the complex configuration of rotating, interacting bluff-body shapes (see **Fig. 1**) producing unsteady turbulent wakes. The problem can be broadly divided between parasitic drag and stability/control. Pressure drag due to separation over the bluff body components of the rotor hub are the dominant form of hub drag. The rotor hub is estimated to be responsible for 20% to 30% of the total drag for single-hub aircraft (Keys & Wiesner, 1975; Sheehy, 1977). Since the empennage/tail (i.e. control surface) is caught in the rotor hub wake, the periodic turbulent structures impacting the control surface have a negative effect on the stability and control in pitch and yaw as well as the helicopter's structural safety (Roesch & Dequin, 1985). In addition, tail shake (i.e. structural vibrations associated with the wake impinging on the tail) can be amplified if the wake has frequency content near the natural frequencies of the empennage assembly.

Despite the hub's importance to vehicle performance, quantitative prediction of rotor hub drag and the resulting wake remains elusive, leading to costly delays in the helicopter design process. Industry has used a variety of computational fluid dynamics (CFD) solvers to predict the magnitude of the steady (average) drag of non-rotating hubs within an accuracy of ~10% (Bridgeman & Lancaster, 2010; Dombroski & Egolf, 2012). However, such solvers are unable to match the unsteady drag harmonics observed experimentally. Academic efforts have been more successful using generic hub geometry to identify dominate frequency content, but additional work was required to accurately predict the corresponding amplitudes (Shenoy et al., 2013). Validation of computational modeling of unsteady wakes rely on comparisons with experimental "long-age" wake studies. Here "long-age" indicates studies that include the wake beyond the location corresponding to the fin and empennage assembly of a helicopter, which is nominally 7 hub radii downstream of the hub. Key "long-age" wake studies are reviewed below, but Reich et al. (2016) provides a more comprehensive review of previous work on helicopter rotor hub drag, wake flow physics and the impact on aircraft stability.

Roesch & Dequin (1985) performed a wind tunnel study with a $1/7^{th}$ scale, four-rotor bladeless helicopter rotor hub (without rotor blades) mounted on a scaled fuselage (AS355 Écureuil 2, Aérospatiale). The hub diameter based Reynolds number ($Re_D$) was varied from $10^5$ to $\sim 4\times 10^5$ ($\sim 1/10^{th}$ of full-scale) with the advance ratio varied from 0.1 to 0.4. Flow visualization showed large coherent turbulent structures propagating downstream, and hot film anemometry measured the velocity fluctuations at a single location upstream of the empennage. The dominant spectral content occurred at the blade passage frequency (i.e. four-per-hub-revolution, '4/rev'), which was associated with the rotor shank geometry. In addition, a prominent (though slightly weaker) 2/rev fluctuation was observed, which was attributed to the scissor link geometry since it was symmetric about 180°. Berry (1997) expanded upon this earlier work using laser Doppler velocimetry (LDV) to measure the turbulent wake. The $1/5^{th}$ scale four-bladed, two-scissor links hub model (US Army 2-Meter Rotor Test System; Phelps & Berry, 1987) was mounted on a generic helicopter fuselage in a wind tunnel and included the rotor blades. The advance ratio was fixed at 0.23 and $Re_D = 1.1 \times 10^6$. An unexpectedly strong 2/rev component was observed in the near wake (less than half a rotor diameter downstream), which was eventually exceeded by the 4/rev component further downstream. While no detailed explanation was offered, the 2/rev content was conjectured to be the product of the interaction between the hub and fuselage based on the location of the turbulent structures. While both Roesch & Dequin (1985) and Berry (1997) observed 2/rev content (though with contradictory explanations), neither study acquired data sufficiently far downstream to correspond to the empennage (tail) assembly location.

More recently, the importance of high Reynolds number testing to produce accurate long-age wake predictions has become more apparent (Shenoy et al., 2013). Consequently, most experimental testing has moved to water tunnels, where higher Reynolds numbers are readily achieved as the kinematic viscosity of water is ~1/15$^{th}$ that of air. The use of water tunnels for rotor hub testing was justified since the maximum local Mach number for the rotor hub of a large helicopter at cruise speeds is ~0.2. Reich et al. (2014a) performed experiments on a 1:4.25 scale model of a de-featured, four-rotor bladeless commercial helicopter hub in the Garfield Thomas Water Tunnel (GTWT). A schematic of the test model is provided in **Fig. 1**, which included upper and lower spider arms, main hub arms (blade shanks), swashplate, vertically angled scissors (set ~30° out-of-phase with the blade shanks) and tilted 5° to simulate forward flight. Unlike the previous studies, the model was not mounted to a helicopter fuselage, but the rotor shaft did have a NACA 0025 fairing. Testing was performed at an advance ratio of 0.2 and $Re_D$ of ~2.5×10$^6$ or ~4.9×10$^6$. Phase-averaged particle-image-velocimetry (PIV) and LDV confirmed the 2/rev and 4/rev content in the near- (~2 hub radii downstream) and long-age (~7 hub radii downstream) wake locations. In the long-age wake, the 2/rev harmonic was stronger than the 4/rev in all vertical locations, except within the rotor blade plane. Since no fuselage geometry was included, it was concluded that both the 2/rev and 4/rev turbulent structures were due to vortex shedding from the 2-per-hub (scissor links) and 4-per-hub (rotor blade shanks) features, respectively, and not a complex fuselage-pylon-wake interaction. In addition, a 6/rev fluctuation was observed at all streamwise locations in the PIV, LDV and drag measurements. Unlike the 2/rev and 4/rev content, the rotor hub had no direct geometric counterpart for the 6/rev. These results were compared with a computational study of a generic four-rotor bladeless hub (Reich et al., 2014b).

The results from Reich et al. (2014a), combined with Shenoy et al. (2013) and Smith & Shenoy (2014), motivated a significant experimental-computational collaboration that has rapidly advanced the subject in recent years. This is an ongoing effort that has included the First Rotor Hub Flow Prediction Workshop (Schmitz et al., 2017; Potsdam et al., 2017; Coder & Foster, 2017; Coder et al., 2017), additional GTWT tests (Schmitz et al., 2017; Metkowski et al., 2018) and the Second Rotor Hub Flow Prediction Workshop that was held in 2018. Most of the data from the additional GTWT tests have not been published, but initial key points are that full scale $Re_D$ tests were achieved, measurement and analysis procedures required for proper comparison between numerical and experimental results have been identified, a decrease in the drag force uncertainty has improved numerical-experimental correlations and a more detailed wake characterization has been developed (Schmitz et al., 2017).

This work and related efforts have resulted in some significant advancements in numerical efforts related to improvements of the fluid/structure interactions (Quon et al., 2012; Jacobson & Smith, 2018) and reducing computational costs with physics-based reduced order modeling (Prosser & Smith, 2015; Koukpaizan et al. 2018a, 2018b). In addition, this collaboration has influenced additional lower Reynolds number water tunnel experiments (Reich et al., 2015; 2017;

2018) using a 1:17 scaled model similar to that of Reich et al. (2014a). These tests used a variety of measurement techniques (phase-averaged 2D PIV, stereo-PIV, drag force sensors and oil-paint visualization) and varied the advance ratio (0.2-0.6), Reynolds numbers ($10^6 \leq Re_D \leq 2.8 \times 10^6$) and model configuration. The key findings from these experimental studies were that the advance ratio does not impact the average drag but does impact the unsteady effects, that pitch links play an important role in the interactional aerodynamics of the unsteady wake, unsteady wake content is Reynolds number dependent, that the 6/rev structure is strongest on the retreating side and that the phase-averaged wake structures favor the retreating side.

All these previous studies have focused on geometric configurations that closely resemble the hub (and sometimes fuselage) geometry of specific commercial helicopters. While this is not surprising given the applied nature of most rotorcraft studies, it has limited the scope of conclusions drawn from them. This is apparent from early studies only being able to conjecture at the cause for specific wake structures. In addition, the complexity of commercial hub geometries force CFD modeling to dedicate high grid resolution at the hub. For example, Dombroski & Egolf (2012) used ~8.2 million of a total of 15 million cells to model the surface and boundary layer relation of a commercial helicopter rotor hub. Thus, performing experiments and computational validation with specific commercial geometry obscures the underlying flow physics responsible for a given flow structure and limits the ability to test CFD modeling accuracy. Consequently, the current work aims to provide a long-age wake characterization within a water tunnel behind a geometrically simple hub composed of bluff bodies with canonical profiles, which can provide more fundamental insights into the wake structure. Specifically, the current study provides insights into the 'mystery' of the 6/rev harmonic, as it was described in Schmitz et al. (2017). While most studies support the view that it is the product of a nonlinear interaction between the 2/rev and 4/rev structures (Schmitz et al., 2017; Potsdam et al., 2017), there have also been alternative hypotheses (e.g. Strouhal shedding from a hub component) proposed within the literature (Reich et al., 2014a, 2018). The current work provides insight into this problem by only including the main rotor arms and scissors link. The relative phase between the main arms and scissors link was varied to assess if the 6/rev can be modified by changes to the main rotor and/or scissors. In addition, the model was tested at 0° angle-of-attack, which greatly simplifies the projected area from the rotor hub.

## 2 Experimental Facility and Methods

### 2.1 Test facility

Testing was performed in the Oklahoma State University 6-inch low-turbulence, recirculating water tunnel (Elbing et al., 2018). The test section had acrylic walls for optical access and measured 1.1 m long with a 152 mm (6 inch) square cross-section. A maximum empty test section speed of 10 m/s was achieved with a horizontal split case centrifugal pump (S10B12A-4, Patterson) that was powered by a 112 kW (150 hp) motor (MP44G3909, Baldor). Flow conditioning with a tandem configuration of honeycombs and settling-chambers, an 8.5:1 area contraction and gradual expansion in diffuser sections resulted in an inlet turbulence level < 0.3% and negligible mean shear within the test-section core. Additional details on the design and characterization of the facility is available in Daniel (2014) and Farsiani et al. (2016).

### 2.2 Test models

Two rotating test models were designed with two objectives; (i) simplify the rotor hub geometry as much as possible without losing the prominent vortex shedding behavior and (ii) achieve a Reynolds number approaching full scale (i.e. within the range of the previous studies). For simplicity, the test models (schematically shown in **Fig. 2**) consisted of only a rotor mast/shaft (1/rev) with rotor blade shank arms (4/rev) and the scissor link arm (2/rev). The models were identical aside from the phase orientation between the rotor blade shank arms and the scissor arms, with one in-phase (0°) and the other 45° out-of-phase (see **Fig. 2**). Other simplifications included making the scissor links parallel with the plane of the rotor blade shanks and testing at 0° angle of attack (typically, testing is performed at a 5° inclination into the flow to simulate forward flight).

The two canonical rotor hub models were fabricated from aluminum (Al-6061) with the rotor blade shank press-fit and welded to the rotor-shaft/scissor-link section. The rotor blade shank arms had a tip-to-tip diameter of 76.2 mm and a 10.2 mm (chord) × 5.1 mm (thick) rectangular cross-section with sharp corners. The rectangular profile with a chord-to-thickness ratio of 2:1 was selected to match established canonical profile results (Delany & Sorensen, 1953). A 7.9 mm radius fillet was used between the rotor blade shank arms. The tip-to-tip diameter of the rotor blade shank arms was fixed at half the test section width to mitigate wall effects, and the other parameters were scaled to be nominally consistent with the geometric ratios from Reich et al. (2014a). The scissor link arm had a tip-to-tip diameter of 25.4 mm and a 10.2 mm (chord) × 2.5 mm (thick) rectangular cross-section with sharp corners. The chord-to-thickness ratio of the rectangular profile was increased to 4:1 for rigidity and manufacturability. The center-to-center vertical spacing between the rotor blade shank arms and the scissor links was 11.4 mm, and the overall vertical distance (top of rotor shank arms to bottom of the rotor shaft that was inserted into the fairing) was 19.1 mm.

A pair of 3D-printed fairings (vertical for rotor shaft and horizontal for tunnel wall, see **Fig. 2**) were used to minimize the wake blockage from the mounting support structure. The vertical fairing had a rectangular planform with a NACA 0015 cross-section. The top of the vertical fairing that mates with the exposed section of the 12.7 mm diameter rotor shaft was flat and parallel with the streamwise flow similar to that used in Reich et al. (2014a). The tunnel wall fairing had a 1.5:1 elliptical leading edge, a flat section for hardware attachment and a linearly decreasing trailing edge with a height-to-length ratio of 13 (>10 is recommended to prevent separation on blister-type fairings; Hoerner, 1965). The coordinate system used for the current work has the $x$-axis aligned in the streamwise direction, $z$-axis aligned with the rotor hub axis of rotation (positive direction points from the scissors to the rotor shank arms) and the $y$-axis completing a right-handed coordinate system. The coordinate origin is located on the rotor hub axis of rotation at the center of the main rotor hub arms.

### *2.3  Instrumentation*

The primary measurement used to characterize the far wake was 2D phase-averaged PIV. Phase-averaged PIV (2D and stereo-) is a standard approach for rotor-wake measurements (2D – Reich et al., 2014a, 2017; stereo – Raffel et al., 2004; Raghav & Komerath, 2015; Reich et al., 2014a, 2018). Note that given the strongly three-dimensional flow-field, stereo-PIV is generally preferred to mitigate the influence of out-of-plane particle motions. However, for the current arrangement 2D PIV was used since it could produce better accuracy of the streamwise velocity component. The errors associated with the out-of-plane particle motion were minimized by using a thin laser sheet and limiting the analysis to the center 60% of the FOV. The PIV system was operated in double-frame, double-pulse mode and phased locked with the rotor hub rotation to produced phase-averaged velocity fields. The image plane was aligned with the rotor shaft ($y = 0$; $-60$ mm $< z < +40$ mm) and centered at $x = 270$ mm (~7 hub radii downstream) as illustrated in **Fig. 3**. Note that a key recommendation from the recent experimental-computational rotor hub research was for experimental data to provide offsets to each side of the PIV plane to quantify variability of the data prior to comparison with computational results. Unfortunately, the current data was acquired prior to these findings so only a single plane was measured. Any future testing should measure the offset planes, and it is recommended that comparisons with the current data should be performed with a sensitivity analysis of the computational results. The image plane was illuminated with a thin 532 nm laser sheet formed with an Nd:YAG laser (Gemini 200, New Wave Research). The flow was flooded with 18 µm diameter hollow glass spheres (iM30K, 3M) to scatter the laser light. The illuminated image plane was recorded with a sCMOS camera (Imager, LaVision) with a resolution of 2560×2160 pixels. The final field-of-view (FOV) was 120 mm × 100 mm, which was achieved with a 60 mm diameter, f/2.8D lens (AF Micro-NIKKOR, Nikon). The images were spatially calibrated with a 58 mm square calibration plate (Type 058-5, LaVision). The velocity vector-fields were computed using standard cross-correlation methods (DaVis 8.2.3, LaVision) with a final interrogation window of 32×32 pixels with 50% overlap, which had a nominal vector spacing of 0.8 mm. For each test configuration (model), a minimum of 2200 total image pairs (vector-fields) were acquired with a nominal phase spacing of 15° (100 realizations per phase).

The PIV uncertainty was quantified following the work of Wieneke (2015), which has been incorporated within the commercial PIV processing software used to process the current results (DaVis 8.2.3, LaVision). This approach uses the computed displacement to transform the raw images to an equivalent time (typically both images are transformed by half the displacement). If properly converged the correlation between these images are at a maximum, and in the absence of noise the correlations would decrease by an equal amount when slightly shifted (~1 pixel) away from the optimized displacement. The asymmetry between these shifted correlations are directly related to the level of noise within the image. Most PIV algorithms correct this asymmetry, which results in an erroneous measured displacement for noisy data. Wieneke (2015) showed that the fluctuations of these corrections can quantify the impact of image noise, including out-of-plane particle motion, on the velocity measurement. This is particularly important for the current study since the rotor wake will produce significant out-of-plane particle motion within the image plane. This analysis resulted in a nominal uncertainty in the current study of ±0.15 m/s (~0.3 pixels), which is larger than the standard 0.1 pixels (nominal limit of correlation algorithms, which does not account for noise) due in part to the three-dimensionality of the flow.

The external trigger for the phase-averaging was produced with a custom-built Hall effect sensor. It used a programmable microcontroller (Uno R3, Elegoo), a Hall effect switch (A3144, Allegro Microsystems) and a magnet rigidly mounted on the rotor shaft. The microcontroller supplied excitation, provided a digital output (for determining hub rotation frequency and phase via time lag from reference position) and a trigger signal for the PIV acquisition system with a desired phase lag. While the exact phase orientation was known to within ±0.5°, during acquisition minor deviations between the actual rotation frequency and the target frequency resulted in slight deviations from the desired 15° increments. Consequently, the in-phase model had 22 phases per revolution (16.3° increments, as opposed to the intended increment of 15°).

The Hall effect sensor output was recorded with the tunnel operation conditions (temperature, static pressure and pump frequency) via a data acquisition card (USB-6218-BNC, National Instruments) and commercial data acquisition software (LabView15.0.1, National Instruments). The tunnel static pressure was measured with a differential pressure transducer (PX2300-50DI, Omega) mounted upstream of the contraction inlet and aligned vertically with the test section centerline. The corresponding test section static pressure was determined accounting for flow acceleration due to the area contraction. The water temperature was measured within ±0.1 °C with a T-type thermocouple (TC-T-1/4NPT-U-72, Omega) located upstream of the test section contraction. The pump motor frequency was manually controlled via a variable frequency drive (EQ7-4150, Teco), which had a digital display as well as an analog output.

### *2.4 Test conditions*

As previously mentioned, experiments were performed using two models, scissor arms (i) in-phase and (ii) 45° out-of-phase with the rotor arms. All testing was performed at a freestream speed ($U_\infty$) at the test section inlet of 9.9 m/s. The models were rotated at 30 Hz to set the advance ratio ($\mu = U_\infty/\omega R$, where $\omega$ is the shaft angular velocity and $R$ is the rotor blade radius) at 0.2, which is consistent with previous studies (Roesch & Dequin, 1985; Phelps & Berry, 1987; Reich et al., 2014a, 2015, 2017, 2018). Note that ratios were being scaled to be consistent with Reich et al. (2014a) (i.e. Sikorsky S-92 helicopter), which the hub radius of the S-92 is ~14% of the rotor blade radius. The tunnel was pressurized to 276 kPa to suppress cavitation at the rotor blade shank arm tips. The nominal water temperature during testing was 20 °C, which has a corresponding kinematic viscosity ($v$) and density of $1.0 \times 10^{-6}$ m$^2$/s and 998 kg/m$^3$, respectively.

The projected area in the streamwise direction for the rotor hub model, vertical fairing (NACA 0015) and tunnel wall fairing was 598 mm$^2$ (maximum), 494 mm$^2$ and 838 mm$^2$, respectively. This produces a solid blockage ratio at the hub of 8.3%, which increases the freestream speed at the hub from 9.9 m/s to 10.8 m/s. Note this is comparable to Reich et al. (2018) that had a 7% blockage that was assumed negligible. With the corrected freestream speed and the hub model diameter ($D_h$) of 76.2 mm, the corresponding hub-diameter-based Reynolds number ($Re_D = U_\infty D_h/\nu$) was $8.2\times10^5$. This corresponds to ~1/3rd scale or ~1/9th scale relative to a small (e.g. Robinson R44) or large (e.g. Sikorsky S-92), respectively. While the solid blockage was significant, the wake blockage was much less since the fairings were designed to have minimal wake contributions even though they constitute most of the solid blockage. Following the analysis of Maskell (1963) with a bluff body constant of 2.5 and a nominal drag coefficient of 0.9 (Schmitz et al., 2017), the current rotor with a blockage of 2.6% produces a 5.8% increase in the dynamic pressure. This corresponds to a 2.9% increase in the freestream speed, which is a conservative estimate given the current model should have a lower drag than Schmitz et al. (2017). This is also consistent with the observation that the downstream wake freestream speed only had ~1% increase relative to the inlet condition. Since the difference between these estimates is within the uncertainty of the velocity measurement, no corrections for wake blockage have been applied. More details about the experimental setup and model design are provided in Petrin (2017).

Table 1 provides an overview of the local Reynolds number ($Re = U_{rel}L/\nu$) and Strouhal number ($St = f_s t/U_{rel}$) for individual components, where $U_{rel}$ ($=U_\infty \pm \omega r$) is the local relative velocity, $r$ is the radial distance from the axis of rotation, $L$ is the characteristic length, $f_s$ is the shedding frequency and $t$ is the thickness (or diameter). Here the Reynolds number characteristic length is either the overall diameter (hub diameter, scissor diameter and rotor shaft) or the chord length of the rectangular profile (rotor blade shank and scissor link). Most of these shedding frequencies are relatively large (> 300 Hz = 10/rev), and the remaining components consist of non-integer numbers of the rotation frequency. Though the main rotor shank at the tip on the advancing side does come close to having a shedding frequency of 2/rev (60 Hz).

## 3 Results

The phase-averaged results for both velocity components (streamwise and vertical) on the in-phase model are stitched together in **Fig. 4** to illustrate the general orientation and phase variation within a hub revolution. The streamwise spatial distribution of vectors was converted to a relative phase position assuming Taylor's frozen turbulence hypothesis (similar to Reich et al., 2014a, 2017), which is valid given that the streamwise velocity fluctuations were less than 5% of the free-stream speed. This assumes the mean flow dominates advection and lag times per vector spacing were estimated by dividing the vector spacing ($dx$) by the average convection velocity for a given $z$-location. These lag times relative to a given reference position allowed for a nominal phase value to be defined for each vector. The $z$-axis is scaled with vertical distance between the rotor blade shanks and the scissor links ($h = 11.4$ mm). The 4/rev oscillations (i.e. 4 cycles within this phase-averaged hub revolution) are seen in both the streamwise and vertical velocity. That is, over the 360° phase-trace four peak-to-peak periods are observed (particularly in the vertical component) with the vertical peaks located near 30°, 120°, 210°, and 300°. The 2/rev content is less noticeable, especially in the streamwise component. However, focusing on the higher speed contributions in the vertical velocity, the second and fourth (from the left) of the aforementioned structures are weaker relative to the first and third structures. Note that a nearly identical image was produced for the in-phase model.

Vertical profiles of the mean streamwise velocity ($U$) at $x = 270$ mm (7 hub radii downstream) for both models are shown in **Fig. 5**. The wake region nominally spans from $-4 < z/h < 2$. Above the wake ($z/h > 2$) both models have a relatively flat profile corresponding to the freestream speed ($U_\infty = 9.95$ m/s). Below the wake ($z/h < -4$), the profiles flatten out at a speed below the freestream due to the shaft fairing wake before decreasing further at the bottom of the FOV due to the developing boundary layer on the tunnel wall. The largest velocity deficit occurs slightly below the scissor link height ($z/h \approx$ -1.2) for both models, which indicates that both wakes were deflected downward (towards the fairing) by equal amounts. This is consistent with Reich et al. (2018), which shows a downwash along their hub model centerline except at the swashplate where flow deflects upward from the 5° angle-of-attack (current model does not have a swashplate and is at 0° angle-of-attack). In addition, the in-phase model wake has a wider wake deficit than the out-of-phase model, which indicates that the losses are larger for the in-phase model.

The vertical profiles of the root-mean-square velocity (i.e. standard deviation at each vertical position) were examined (not shown), which both models had a peak fluctuating velocity < 5% of the mean streamwise velocity. Thus, streamwise velocity fluctuations ($u'$) are small relative to $U_\infty$, which justifies the previous use of Taylor's frozen turbulence hypothesis. The in-phase model peak occurred lower ($z/h$ = -2.3) than the out-of-phase model ($z/h$ = -1.8). The fluctuating velocity profiles for both models are examined in greater detail in **Fig. 6**. Here the mean squared fluctuating velocity components ($u'^2, w'^2, u'w'$) are scaled with the square of the velocity deficit ($U_S = U_\infty - U_{min}$), where $U_{min}$ is the minimum mean velocity within the wake ($U_S$ = 1.5 m/s). Included are dashed lines corresponding to the uncertainty bands, which were determined following the analysis of Schiacchitano & Wieneke (2016) that used the standard deviation of the uncertainty determined from the Wieneke (2015) analysis. Similar to the mean profiles, these profiles show that the wake of the in-phase model was wider than that of the out-of-phase model. In addition, the in-phase model has larger peaks for all components with the exception of the secondary peak in the $u'^2$ profile at the main rotor height ($z$ = 0). Taken together, these profiles bear a superficial resemblance to the self-similar axisymmetric wake described in Pope (2000) with the wake center near the scissor height ($z/h \sim -1$).

Phase traces at a given height ($z$) can be extracted from the phase-averaged contour plots in **Fig. 4**, which an example of one such phase trace at $z/h = -2$ is provided in **Fig. 7** with dashed lines indicating the uncertainty from the Wieneke (2015) analysis for each data point. Note that the streamwise component ($u$) of the in-phase model is consistently lower than the out-of-phase model, which is consistent with the wider spreading of the wake (see **Fig. 5**). It is also important to note that the vertical component of the out-of-phase model has asymmetric peaks between peaks that are similar to the in-phase model. This indicates that there is coherent higher frequency content within the wake. Spectral analysis was performed by transforming the phase traces to a spatial distribution (or temporal) by applying Taylor's frozen turbulence hypothesis. The wavenumber-domain spectra, $S_{uu}(k)$, of the wake was produced from the fast Fourier transform (FFT) of the traces.

The spectral levels at frequencies corresponding to 1 to 10 cycles per hub revolution were extracted from the resulting power spectra. As will be shown subsequently, the three frequencies that contained the most energy and will be the focus of the current discussion are the 2/rev, 4/rev and 6/rev. **Fig. 8** shows the vertical distribution of the spectral levels of the $i^{th}$ per revolution contributions in the streamwise ($u_i$) and vertical ($w_i$) velocity, which is scaled with $U_\infty$. The uncertainty bands (i.e. dashed lines) were estimated using a Monte Carlo simulation of 100,000 phase traces produced from the mean phase trace in **Fig. 7** assuming each data point was normally distributed about the mean value with a standard deviation equal to the phase trace uncertainty. The standard deviation of the peak amplitude at the frequency of interest was used to define the uncertainty bands. The distribution of the 4/rev content is very similar between the in-phase and out-of-phase models with the out-of-phase model having slightly higher peak values. For both models, the 4/rev streamwise component has a bimodal distribution with peaks close to the main rotor arm height ($z/h = 0$) and below the scissors close to the fairing height ($z/h \sim -2$) with a significant drop in spectral level near the scissors elevation ($z/h = -1$). Conversely, the 2/rev content has a peak close to the scissors location though the in-phase model peak appears to be pushed further down ($z/h \sim -2$). The sharp drop in the 4/rev aligned with the 2/rev peak indicates a potential complex interaction between these two structures. The 2/rev component of the vertical velocity is weak for both models with a broad peak approximately centered on the scissors height. The out-of-phase model has stronger 6/rev spectral levels with asymmetric peaks pushed to the lower side of the wake (i.e. towards the scissors location). It should be noted that there does appear to be weaker peaks in the 6/rev distribution for the in-phase model that nominally correspond to those stronger peaks in the out-phase-model, but at different vertical locations.

# 4  Discussion

## *4.1  Dynamic mode decomposition (DMD) analysis*

While time-series analysis has been the backbone of experimental analysis, Taylor's frozen turbulence hypothesis is often leveraged to extract the spatial evolution. However, PIV datasets, such as those generated in this study, offer unique planar spatiotemporal information that can be leveraged to analyze characteristic flow structures and identify instability models if present. Traditionally, identification of coherent structures is accomplished using (spatial) proper orthogonal decomposition (POD) technique (Lumley, 2007; Holmes, 2012; Taira et al., 2017), also known as principal component analysis (PCA) and singular value decomposition (SVD). Spatial

POD modes represent coherent structures that occur with a high degree of probability within the dataset, as they are essentially eigenvectors of the spatial two-point correlation tensor sorted in terms of energy content. Consequently, each spatial POD mode may contain multiple temporal frequencies that are bunched together based on energetic similarity. Often times, such as in this case of helicopter wake dynamics, it is useful to identify coherent structures associated with a characteristic temporal frequency. This requires methods such as spectral POD (Towne et al., 2018), a time-spectral analogue of the spatial POD or linear system theory based dynamic mode decomposition (DMD).

DMD is an algorithm (Schmid, 2010; Rowley et al., 2009; Rowley & Dawson, 2017) for carrying out spectral analysis of time-resolved data. The essence of this approach is the computation of modes through eigendecomposition of the Markov linear transition operator learned using data snapshots. Each of these modes are associated with a specific frequency and growth rate. For a linear system (i.e. a system governed by linear dynamics), the DMD modes are the normal modes. However, in general for a nonlinear system, these modes are projections of the eigenmodes of the Koopman operator (adjoint of the more well-known Perron-Frobenius operator) to the space of the full flow state (Koopman, 1931; Mezić, 2005; Rowley et al., 2009). Consequently, the DMD or Koopman modes with intrinsic temporal signatures are non-orthogonal and therefore, less compact than POD modes for low-dimensional representations. In spite of this, they are attractive tools that provide physically and dynamically meaningful information about the flow. In this study, the DMD algorithm is leveraged to identify the spatial structure of the dynamics within the helicopter hub wake experimental data, specifically focusing on the time scales associated with the 2/rev, 4/rev and 6/rev content.

The key steps constituting the DMD algorithm are presented here for the benefit of the reader. Starting with time-separated data organized into snapshot (i.e. instantaneous vector field) pairs denoted by $\boldsymbol{X}_T = [\boldsymbol{x}_T, \boldsymbol{x}_{T+\Delta T}, \ldots, \boldsymbol{x}_{T+(M-1)\Delta T}]$ and $\boldsymbol{X}_{T+\Delta T} = [\boldsymbol{x}_{T+\Delta T}, \boldsymbol{x}_{T+2\Delta T}, \ldots, \boldsymbol{x}_{T+M\Delta T}]$, where $\boldsymbol{X}_T, \boldsymbol{X}_{T+\Delta T} \in \mathbb{R}^{N \times M}$, $N$ is the dimension of the measurement data, $M$ is the total number of data snapshots (i.e. 2200 or 2400) and $\Delta T$ is the separation time between realizations (1.5 μsec), one characterizes the temporal evolution of the nonlinear fluid flow system $(\boldsymbol{X}_{T+\Delta T} = \boldsymbol{F}(\boldsymbol{X}_T))$ using a Markov linear approximation of the dynamics governed by $A$ in a feature space (also called a Koopman operator; Mezić, 2005, 2013) generated by a map $\boldsymbol{g}$ such that,

$$\boldsymbol{g}(\boldsymbol{X}_{T+\Delta T}) = A\boldsymbol{g}(\boldsymbol{X}_T). \tag{1}$$

In the DMD framework, the feature map $(\boldsymbol{g})$ is approximated as the linear functions of full flow state using the transposed left singular vector of the data matrix $D$ (i.e. $\boldsymbol{g}(\boldsymbol{X}) = D^T\boldsymbol{X}$), as obtained from the SVD problem,

$$\boldsymbol{X}_T = D\Sigma W^T. \tag{2}$$

The left singular vectors $D$ are also the POD modes of the data if the mean is removed. Using Eq. (1) and (2), the Markov linear model is transformed as

$$AD^T\boldsymbol{X}_T = A\Sigma W^T = D^T\boldsymbol{X}_{T+\Delta T}, \tag{3}$$

which on further simplification allows the Koopman operator ($A$) to be approximated as

$$A = D^T\boldsymbol{X}_{T+\Delta T}W\Sigma^{-1}. \tag{4}$$

Here $\Sigma W^T$ represents the features, $D^T\boldsymbol{X}_T = \boldsymbol{g}(\boldsymbol{X}_T)$. The Koopman or DMD modes are the projections of the eigenvectors of $A$ onto the space containing the full flow state using the left singular vectors. Thus, if $\phi$ represents the eigenvector of $A$, the DMD mode is given by $\zeta = D\phi$. The growth rate ($\alpha$) and frequency ($f$) for each of the DMD modes are computed from the eigenvalues $(\lambda = \lambda_R + i\lambda_I)$ as $\alpha = \ln\left(\sqrt{\lambda_R^2 + \lambda_I^2}\right)/\Delta T$ and $f = \tan^{-1}(\lambda_I/\lambda_R)/(2\pi\Delta T)$, where $\lambda_R$ and $\lambda_I$ are the real and imaginary parts of the eigenvalues, respectively.

The phase trace data in **Fig. 7** was composed of at least 2200 snapshots (100 per phase). Each phase increment ($d\theta$) can be interpreted as the time step $\Delta T$ (= $d\theta/2\pi f_{hub}$, where $d\theta$ is in radians and $f_{hub}$ is the hub rotation rate in Hz), which allows the 2200 to 2400 snapshots each with phase information to be turned into an equivalent time-trace. If anything, this is likely to impact higher frequency content, but the current analysis focuses on the relatively low frequency components (i.e. < 10/rev). An example of the DMD results (real, imaginary and magnitude) from the in-phase model for the 2/rev, 4/rev and 6/rev contributions of the streamwise velocity component are shown in **Fig. 9**. The consistency with the DMD analysis and the previous spectral analysis is apparent by comparing the 4/rev magnitude with the results shown in **Fig. 8**. Here both show that the 4/rev content within the wake has a minimum near the height of the scissors and peak values on each side ($z/h \approx 0$ and -2) of this minimum. The real and imaginary parts of the DMD analysis show the coherent structures associated with the 2/rev and 4/rev frequencies. The 2/rev structures are slightly inclined and fill the wake region, while the 4/rev structures are split nominally at the scissors height with the structures above and below out-of-phase with each other. Conversely, the 6/rev contribution remains relatively incoherent with only small, weak structures in an unstructured distribution. The streamwise results for the out-of-phase model are similar to **Fig. 9**, though the 2/rev structures are not as inclined and there is higher coherence observed in the 6/rev.

**Fig. 10** provides a comparison of in-phase and out-phase model DMD results (imaginary only) for the 2/rev, 4/rev and 6/rev contributions. Here it is apparent that the structure of the 2/rev and 4/rev structures are similar between the two models. However, the 6/rev exhibits much higher coherence with the out-of-phase model. This suggests that the 6/rev structures could be the product of a non-linear interaction between the 2/rev (scissors) and the 4/rev (main rotor arms), which is in contrast to previous conjecture that such content was a Strouhal type shedding. In addition, these results indicate that this effect requires some misalignment between the components. This is particularly intriguing given the findings in Schmitz et al. (2017) that the 6/rev content is consistent with analysis of projected frontal areas. This suggests that the simpler geometry for the current study at 0° angle-of-attack should produce weaker 6/rev content than Reich et al. (2014a), and the in-phase should be weaker than the out-of-phase.

The growth rates and modal energies (normalized by the mean flow mode) are compared between the models in **Fig. 11**. Since all the "growth" rates for the current study are negative, they are more appropriately termed decay rates. For both models, the 2/rev and 4/rev have significantly weaker decay rates relative to the other components. In addition, their modal energies are higher, which means that they carry most of the energy and dissipate slowly. Hence, it is critical that both components be accurately modeled for helicopter design since these structures will persist into the far wake where the helicopter control surface (i.e. tail) is located. For the out-of-phase model, the 6/rev component has a more rapid decay rate and carries less energy relative to the 4/rev, but has a measurable separation (weaker decay rate and higher model energy) from any other frequency component indicating that it should also be considered in the wake analyses.

### *4.2 Comparison with Reich et al. (2014a)*

While additional high-Reynolds number long-age wake surveys (Schmitz et al., 2017; Metkowski et al., 2018) have been performed since Reich et al. (2014a), the wake survey data have not been published. Reich et al. (2017, 2018) have wake survey results, but Reich et al. (2017) primarily examined variations with additional model features (pitch links and beanie fairing), which makes it difficult to compare results due to the changing model complexity. Reich et al. (2018) used stereo-PIV to characterize the wake behind a configuration similar to Reich et al. (2014a). This is convenient for quantifying the expected sensitivity of the image plane location for the current results, even though only contour plots of the fluctuating components were shown. For these reasons the current results are directly compared against Reich et al. (2014a), though the more recent results are incorporated into the comparison and discussion.

**Fig. 12** compares the results from the current work using canonical geometry at zero angle-of-attack to Reich et al. (2014a) that used a defeatured commercial helicopter hub operating at 5° angle-of-attack. The Reich et al. (2014a) model scissors were also out-of-phase (30° or 60° depending on the pair of main rotor arms the angle is referenced from) and not parallel with the main shaft arms (see **Fig. 1**). While Reich et al. (2014a) only reports spectral levels from the vertical ($w$) velocity component, those results are compared with both the streamwise and vertical components of the current study since the Reich et al. (2014a) model was tilted 5°. To summarize the trends from Reich et al. (2014a), the vertical velocity spectra within the far wake (7 hub radii downstream) have 2/rev and 4/rev content of near-equal strength directly behind the rotor main shaft arms. Both weaken moving down from the main rotor arms towards the scissor links, though the 4/rev weakens at a slightly faster rate than the 2/rev. The 6/rev frequency content was relatively weak at the rotor main shaft arms ($z = 0$), but below ($z < 0$) the 6/rev rapidly increases and remains nearly constant below the scissors. These results are consistent with the trends and magnitudes shown in Reich et al. (2018).

Comparison between the current models and Reich et al. (2014a) shows that the 4/rev content is quite similar in magnitudes and trends with the streamwise component. Note that the vertical 4/rev trends are opposite, with Reich et al. (2014a) increasing with increasing $z$ and the current models decreasing with $z$. This is likely related to Reich et al. (2014a) being tilted 5° since those vertical fluctuations so closely follow the current streamwise fluctuations. The 2/rev at scissors ($z/h$ = -1) match between Reich et al. (2014a) and the out-of-phase streamwise components. Otherwise the current models consistently have lower 2/rev amplitudes (streamwise and vertical components) and the trends are opposite with the current models decreasing with increasing $z$. The model and test condition deviations between the two studies include geometry (canonical versus simplified commercial), angle-of-attack (0° versus 5°) and relative phase between rotor arms and scissors (0° or 45° versus 30°/60°). It is difficult to identify the cause for these deviations between the two studies, but the relative phase is unlikely to be the cause since the current study varied that without an apparent change in the trends between the in-phase and out-of-phase models.

All of the models (in-phase, out-of-phase and Reich et al., 2014a) have weak vertical 6/rev fluctuations at the main rotor arms that then increases moving towards the scissors. However, the magnitude of the Reich et al. (2014a) model is significantly larger. More specifically, below the main rotor arms ($z < 0$) Reich et al. (2014a) is the largest, followed by the out-of-phase model and then the in-phase model. This is consistent with both the DMD analysis, which showed that the out-of-phase model had much more coherent 6/rev structures than the in-phase. In addition, these results are consistent with the projected frontal area analysis of Schmitz et al. (2017), which notes that a 6/rev component is produced from the hub frontal area projections due to the complex hub rotor geometry of Reich et al. (2014a) titled at 5°. Specifically, Schmitz et al. (2017) notes that the 6/rev content was produced due to the projected frontal area interactions of the 4/rev blade stubs and spiders with the 2/rev scissors. Thus the expectation for current models based on this analysis is that the 6/rev content would be weaker than Reich et al. (2014a) given that the current models are at 0° angle-of-attack with all horizontal components.

## 5 Conclusions

The current study used phase-averaged PIV to characterize the long-age wake region behind two canonical helicopter hubs, each consisting of a 4/rev main rotor arm and a 2/rev smaller scissors link. The wakes were characterized in terms of their mean and fluctuating velocity profiles as well as the vertical distribution of specific spectral content (2/rev, 4/rev and 6/rev). Consequently, the current work has provided a detailed wake characterization behind a geometrically simple hub composed of bluff bodies with canonical profiles, which can be used for validation of physics-based computational models. Moreover, analysis of the wakes combined with comparison to past studies have produced the following conclusions about the behavior and sensitivity to model configuration:

1) The 2/rev and 4/rev coherent structures are similar in size and orientation between models, which suggests that these structures are not sensitive to the relative phase angle between the scissors and the main arms. Furthermore, DMD analysis showed that most of the energy was contained within these two structures, which also had decay rates that were weaker than any other frequency. Thus accurate prediction of size, strength and decay rate of these structures is critical since they will persist to the helicopter tail.

2) The 6/rev coherent structures are sensitive to the relative phase angle between the scissors and the main arms, which supports the view that the 6/rev structures are produced from a nonlinear interaction between the 2/rev and 4/rev structures. More specifically, DMD analysis showed that the out-of-phase model had 6/rev content that decayed slower and had energy levels above the "background" (frequencies other then 2/rev and 4/rev). Conversely, the 6/rev content for the in-phase model was not significantly different from the background frequencies.

3) Comparison with previous helicopter hub far wake results (Reich et al., 2014a; 2018), showed that the simple, canonical geometry produced similar scaled amplitudes though there were discrepancies in the trends between some of the components. The source of these deviations could not be determined, but are most likely associated with the angle-of-attack difference (0° versus 5°) and/or relative phase between the scissors and main arms (0° or 45° versus 30°/60°).

4) Weaker 6/rev content for the current models relative to the Reich et al. (2014a) is consistent with the projected frontal area analysis from Schmitz et al. (2017), which notes that a 6/rev drag contribution is produced from the interactions of the projected frontal area for the 4/rev (blade stubs and spiders) and 2/rev (spider) features. Since the current models have all horizontal features and 0° angle-of-attack, these complex projected frontal area projections do not exist for the current model.

Thus the current study has characterized the long-age wake behind a simplified helicopter hub composed of bluff bodies with canonical profiles, which has provided fundamental insights into the wake structure and their dependence on the hub geometry and orientation. Specifically, the conjecture from Reich et al. (2014a) that the 2/rev and 4/rev coherent structures in the far wake were produced from the main rotor arms and the scissors link was confirmed by replicating the behavior with simplified geometry that only preserved the basic orientation of these two features. In addition, the 6/rev structures were also definitively shown to be sensitive to the relative angle between the scissor links and rotor arms, which indicates that these are the product of a nonlinear interaction between the 2/rev and 4/rev components as proposed by Schmitz et al. (2017) and Potsdam et al. (2017).

## Acknowledgements

The authors would like to thank Mr. Jacob Niles for 3D printing the fairing, the DML staff that assisted with fabrication of the hub models (Mr. John Gage, Mr. Carson Depew and Mr. Joe Preston), Dr. Andrew Arena for use of equipment, and the MS thesis committee of C.E. Petrin (Drs. Jamey Jacob and Kurt Rouser) for reviewing material. This work was supported in part by B.R. Elbing's Halliburton Faculty Fellowship endowed professorship.

## References

Berry JD (1997) "Unsteady velocity measurements taken behind a model helicopter rotor hub in forward flight," NASA Technical Memorandum, TM-4738, Hampton, VA.

Bridgeman JO & Lancaster GT (2010) "Physics-based analysis methodology for hub drag prediction," *Proceedings of the 66th Annual American Helicopter Society Forum*, 66-2010-000214, Phoenix, AZ (May 11-13).

Coder JG & Foster NF (2017) "Structured, overset simulations for the 1st Rotor Hub Flow Workshop," *Proceedings of the 73rd Annual American Helicopter Society Forum*, 73-2017-0098, Fort Worth, TX (May 9-11).

Coder JG, Cross PA & Smith MJ (2017) "Turbulence modeling strategies for rotor hub flows," *Proceedings of the 73rd Annual American Helicopter Society Forum*, 73-2017-0023, Fort Worth, TX (May 9-11).

Daniel L (2014) "Design and installation of a high Reynolds number recirculating water tunnel," M.S. Thesis, Oklahoma State University, USA.

Delany NK & Sorensen NE (1953) "Low-Speed Drag of Cylinders of Various Shapes," National Advisory Committee for Aeronautics (NACA), Technical Note 3038, Washington, DC.

Dombroski M & Egolf TA (2012) "Drag prediction of two production rotor hub geometries," *Proceedings of the 68th Annual American Helicopter Society Forum*, 68-2012-000269, Fort Worth, TX (May 1-3).

Elbing BR, Daniel L, Farsiani Y & Petrin CE (2018) "Design and validation of a recirculating, high-Reynolds number water tunnel," *Journal of Fluids Engineering*, **140**, 081102-6.

Farsiani Y & Elbing BR (2016) "Characterization of a custom-designed, high-Reynolds number water tunnel," *ASME Fluids Engineering Division Summer Meeting*, FEDSM2016-7866, Washington, DC (July 10-14).

Hoerner, SF (1965) *Fluid Dynamic Drag: Practical Information on Aerodynamic Drag and Hydrodynamic Resistance*, 2nd Ed., Hoerner Fluid Dynamics.

Holmes P, Lumley JL, Berkooz G & Rowley CW (2012) *Turbulence, Coherent Structures, Dynamical Systems and Symmetry*, 2nd Ed., Cambridge University Press, New York.

Jacobson KE & Smith MJ (2018) "Carefree hybrid methodology for rotor hover performance analysis," *Journal of Aircraft*, **55**(1), 52-65.

Keys CN & Wiesner R (1975) "Guidelines for reducing helicopter parasite drag," *Journal of the American Helicopter Society*, **20**(1), 31-40.

Knisely CW (1990) "Strouhal numbers of rectangular cylinders at incidence: A review and new data," *Journal of Fluids and Structures*, **4**, 371-393.

Koopman BO (1931) "Hamiltonian systems and transformation in Hilbert space," *Proceedings of the National Academy of Sciences*, **17**(5) 315-318.

Koukpaizan NK, Wilks AL, Grubb AL & Smith MJ (2018a) "Reduced-order modeling of complex aerodynamic geometries using canonical shapes," *Proceedings of the AIAA Modeling and Simulation Technologies Conference*, 209959, Kissimmee, FL (Jan 8-12).

Koukpaizan NK, Afman JP, Wilks AL & Smith MJ (2018b) "Rapid vehicle aerodynamic modeling for use in early design," *Proceedings of the American Helicopter Society International Technical Meeting*, sm-aeromech_2018_58, San Francisco, CA (Jan 16-19).

Lumley, JL (2007) *Stochastic Tools in Turbulence*, Dover Publications, New York.

Maskell, E.C. (1963) "A theory of the blockage effects on bluff bodes and stalled wings in a closed wind tunnel," Technical Report: ARC Reports and Memoranda R&M 3400, Aeronautical Research Council, London.

Metkowski L, Reich D, Sinding K, Jaffa N & Schmitz S (2018) "Full-scale Reynolds number experiment on interactional aerodynamics between two model rotor hubs and a horizontal stabilizer," *Proceedings of the 74th Annual American Helicopter Society Forum*, 74-2018-0144, Phoenix, AZ (May 14-17).

Mezić I (2005) "Spectral properties of dynamical systems, model reduction and decompositions," *Nonlinear Dynamics*, **41**(1) 309-325.

Mezić I (2013) "Analysis of fluid flows via spectral properties of the Koopman operator," *Annual Review of Fluid Mechanics*, **45**, 357-378.

Monniaux D (2016) "File:Sikorsky S-92 rotor P1230176.jpg," *Wikimedia Commons, the free media repository*, photo from the 2007 Paris Air Show; retrieved February 19, 2018 from https://commons.wikimedia.org/w/index.php?title=File:Sikorsky_S-92_rotor_P1230176.jpg&oldid=192883546.

Norberg C (1993) "Flow around rectangular cylinders: Pressure forces and wake frequencies," *Journal of Wind Engineering and Industrial Aerodynamics*, **49**, 187-196.

Okajima A (1982) "Strouhal number of rectangular cylinders," *Journal of Fluid Mechanics*, **123**, 379-398.

Petrin CE (2017) "Frequency content in the wakes of rotating bluff body helicopter hub models," M.S. Thesis, Oklahoma State University, USA.

Phelps AE & Berry JD (1987) "Description of the U.S. Army small-scale 2-meter rotor test system," NASA Technical Memorandum, TM-87762, Hampton, VA.

Pope SB (2000) *Turbulent Flows*, Cambridge University Press, New York.

Potsdam M, Cross P, Jacobson K, Hill M, Singh R & Smith MJ (2017) "Assessment of CREATE-AV Helios for complex rotating hub wakes," *Proceedings of the 73rd Annual American Helicopter Society Forum*, 73-2017-0225, Fort Worth, TX (May 9-11).

Prosser DT & Smith MJ (2015) "A physics-based, reduced-order aerodynamics model for bluff bodies in unsteady, arbitrary motion," *Journal of the American Helicopter Society*, **60**(3), 032012.

Quon EW, Smith MJ, Whitehous GR & Wachspress D (2012) "Unsteady Reynolds-averaged Navier-Stokes-based hybrid methodologies for rotor-fuselage interaction," *Journal of Aircraft*, **49**(3), 961-965.

Raffel M, Richard H, Ehrenfried K, Van der Wall B, Burley C, Beaumier P, McAlister K & Pengel K (2004) "Recording and evaluation methods of PIV investigations on a helicopter rotor model," *Experiments in Fluids*, **36**, 146-156.

Raghav V & Komerath N (2015) "Advance ratio effects on the flow structure and unsteadiness of the dynamic-stall vortex of a rotating blade in steady forward flight," *Physics of Fluids*, **27**, 027101.

Reich DB, Elbing BR, Berezin CR & Schmitz S (2014a) "Water tunnel flow diagnostics of wake structures downstream of a model helicopter rotor hub," *Journal of the American Helicopter Society*, **59**, 032001.

Reich DB, Shenoy R, Schmitz S & Smith MJ (2014b) "An assessment of the long-age unsteady rotor hub wake physics for empennage analysis," *Proceedings of the 70th Annual American Helicopter Society Forum*, 70-2014-0001, Montréal, Québec (May 20-22).

Reich D, Willits S & Schmitz S (2015) "Effects of Reynolds number and advance ratio on the drag of a model helicopter rotor hub," *Proceedings of the 71$^{st}$ Annual American Helicopter Soceity Forum*, 71-2015-063, Virginia Beach, VA (May 5-7).

Reich D, Shenoy R, Smith M & Schmitz S (2016) "A review of 60 years of rotor hub drag and wake physics: 1954-2014," *Journal of the American Helicopter Society,* **61**(2), 1-17.

Reich D, Willits S & Schmitz S (2017) "Scaling and Configuration Effects on Helicopter Rotor Hub Interactional Aerodynamics," *Journal of Aircraft*, **54**(5), 1692-1704.

Reich D, Sinding K & Schmitz S (2018) "Visualization of a helicopter rotor hub wake" *Experiments in Fluids*, **59**(7):116.

Roesch P & Dequin A-M (1985) "Experimental research on helicopter fuselage and rotor hub wake turbulence," *Journal of the American Helicopter Society*, **30**(1), 43-51.

Rowley CW, Mezić I, Bagheri S, Schlatter P & Henningson DS (2009) "Spectral analysis of nonlinear flows," *Journal of Fluid Mechanics*, **641**, 115-127.

Rowley CW & Dawson STM (2017) "Model reduction for flow analysis and control," *Annual Review of Fluid Mechanics*, **49,** 387-417.

Schiacchitano A & Wieneke B (2016) "PIV uncertainty propagation," *Measurement Science and Technology*, **27**, 084006.

Schmid PJ (2010) "Dynamic mode decomposition of numerical and experimental data," *Journal of Fluid Mechanics* **656**, 5-28.

Schmitz S, Reich D, Smith MJ & Centolanza LR (2017) "First Rotor Hub Flow Prediction Workshop Experimental Data Campaigns and Computational Analyses," *Proceedings of* *the 73rd Annual American Helicopter Society Forum*, 73-2017-0128, Fort Worth, TX (May 9-11).

Sheehy TW (1977) "A general review of helicopter rotor hub drag data," *Journal of the American Helicopter Society*, **22**(2), 2-10.

Shenoy R, Holmes M, Smith MJ & Komerath NM (2013) "Scaling evaluations on the drag of a hub system," *Journal of the American Helicopter Society*, **58**(3), 1-13.

Smith MJ & Shenoy R (2013) "Deconstructing hub drag," Office of Naval Research Final Technical Report, N0001409-1-1019.

Taira K, Brunton SL, Dawson STM, Rowley CW, Colonius T, McKeon BJ, Schmidt OT, Gordeyev S, Theofilis V & Ukeiley LS (2017) "Modal analysis of fluid flows: An overview," *AIAA Journal*, **55**(12), 4013-4041.

Tanaka H & Nagano S (1973) "Study of flow around a rotating circular cylinder," *Bulletin of the Japan Society of Mechanical Engineers*, **16**(92), 234-243.

Towne A, Schmidt OT & Colonius T (2018) "Spectral proper orthogonal decomposition and its relationship to dynamic mode decomposition and resolvent analysis," *Journal of Fluid Mechanics*, **847**, 821-867.

Wieneke B (2015) "PIV uncertainty quantification from correlation statistics," *Measurement Science & Technology*, **26**, 074002.

# Figures and Tables

Table 1. Summary of Strouhal shedding for individual components of the hub models. The scissor diameter and rotor shaft estimates are based on rotating cylinder data. The rotor blade shank and scissor link values are for rectangular cylinders with aspect ratios of 2:1 and 4:1, respectively.

| Component | Side | $L^*$ (mm) | $U_{rel}$ (m/s) | $Re$ (×$10^5$) | $St$ (--) | $f_s$ (Hz) | Source |
|---|---|---|---|---|---|---|---|
| Hub diameter | NA | 76.2 | 10.8 | 8.2 | NA | | |
| Rotor Blade Shank | Advancing | 10.2 | 18.0 | 1.8 | 0.09 | 320 | Knisely (1990) |
| Rotor Blade Shank | Retreating | 10.2 | 3.6 | 0.4 | 0.09 | 64 | Okajima (1982); Knisely (1990) |
| Scissor diameter | NA | 25.4 | 10.8 | 2.7 | 0.20 | 85 | Tanaka & Nagano (1973) |
| Scissor Link | Advancing | 10.2 | 13.2 | 1.3 | 0.14 | 730 | Knisely (1990); Norberg (1993) |
| Scissor Link | Retreating | 10.2 | 8.4 | 0.9 | 0.14 | 460 | Knisely (1990); Norberg (1993) |
| Rotor Shaft | NA | 12.7 | 10.8 | 1.4 | 0.19 | 160 | Tanaka & Nagano (1973) |

*Either diameter (hub, scissor, rotor shaft) or chord length (rotor blade shank, scissor link)

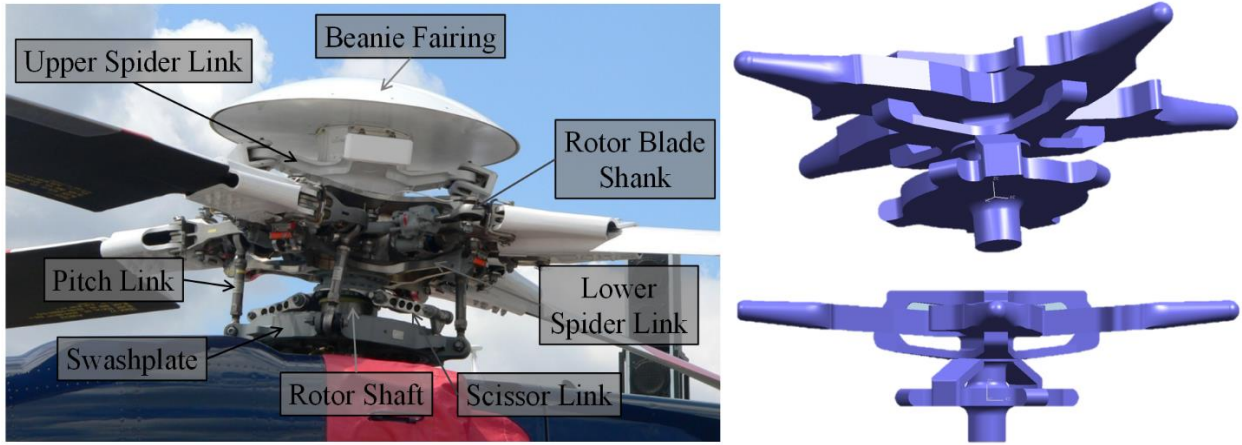

**Fig. 1** (left) Picture of a commercial helicopter rotor hub (S-92, Sikorsky) with the primary components labeled. (right) Schematics of the defeatured model used in Reich et al. (2014a). Left image adapted from Monniaux (2016) under the Creative Commons Attribution 3.0 license. ©David Monniaux

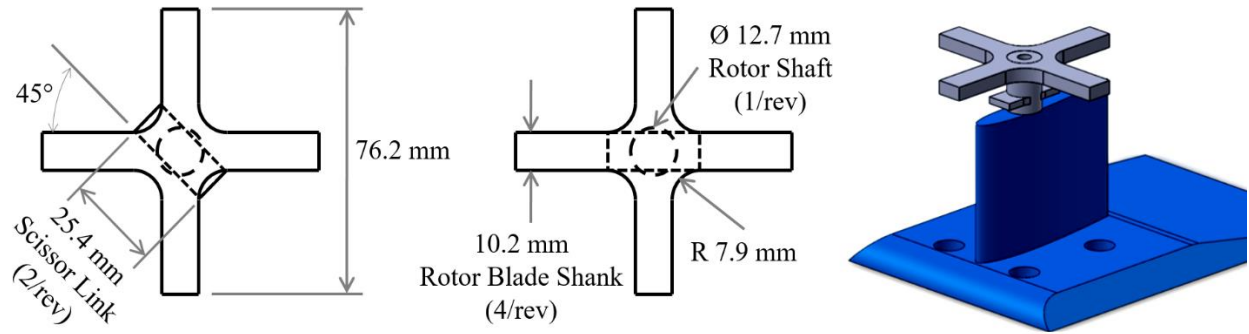

**Fig. 2** Schematics of the test models with the scissor link (left) out-of-phase and (middle) in-phase with the rotor blade shanks. (right) An isometric view of the in-phase model mounted with the fairing for the rotor shaft as well as the fairing at the tunnel wall

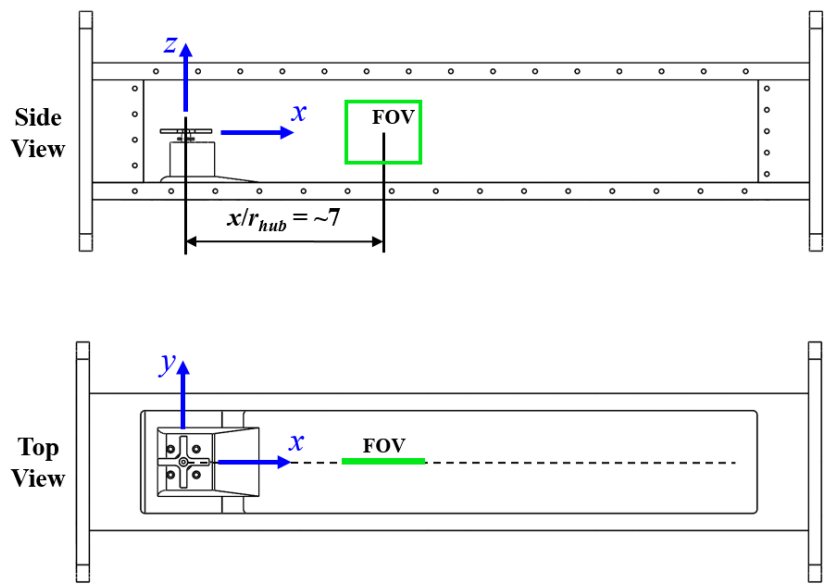

**Fig. 3** Top- and side-view schematics of the test model mounted in the water tunnel test section along with the coordinate system and the nominal location of the PIV FOV

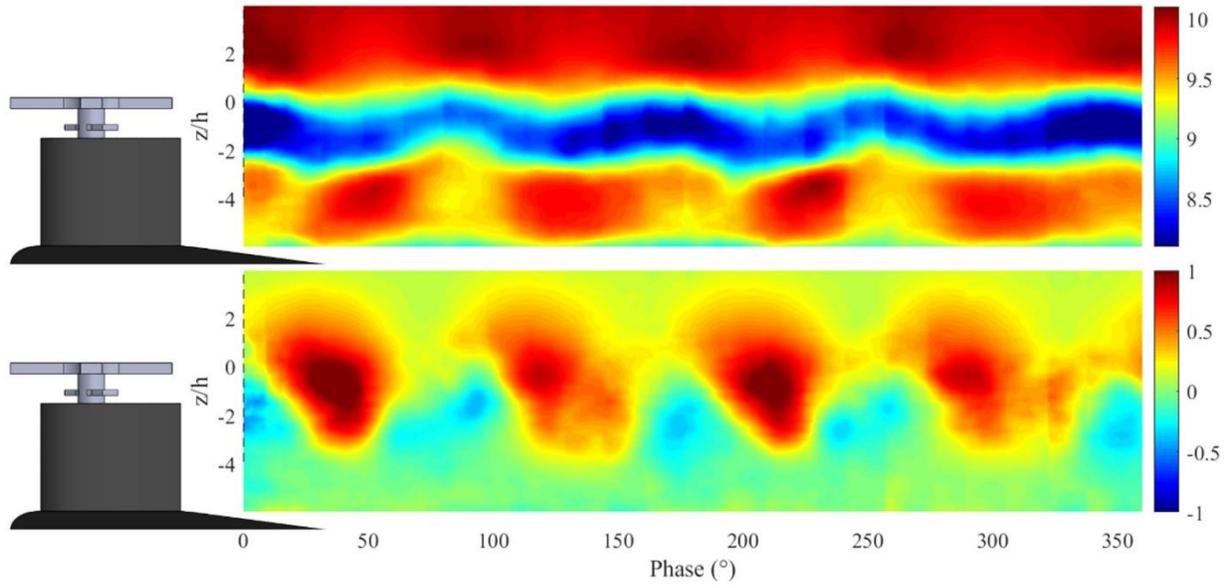

**Fig. 4** Contour maps of the phase-averaged (top) streamwise and (bottom) vertical velocity (m/s) distribution within the far wake of the out-of-phase model. As an orientation/scale reference a schematic of the helicopter hub is provided for each contour map

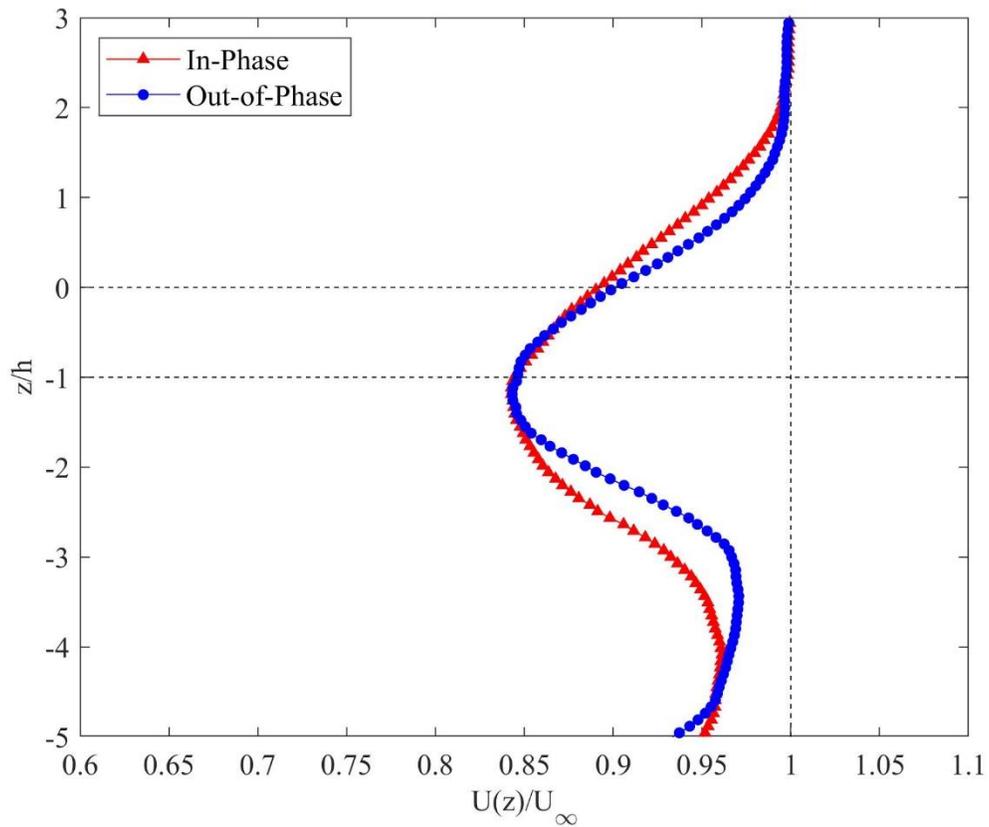

**Fig. 5** Mean streamwise velocity scaled with the local freestream ($U_\infty$ = 9.95 m/s) profiles for both models. Profiles have been extracted from $x$ = 270 mm (7 hub radii downstream of the hub). Vertical distance is scaled with the distance between the rotor arm shanks and the scissors ($h$ = 11.4 mm). Dashed horizontal lines correspond to the center of the rotor arms ($z$ = 0) and scissors ($z$ = -$h$)

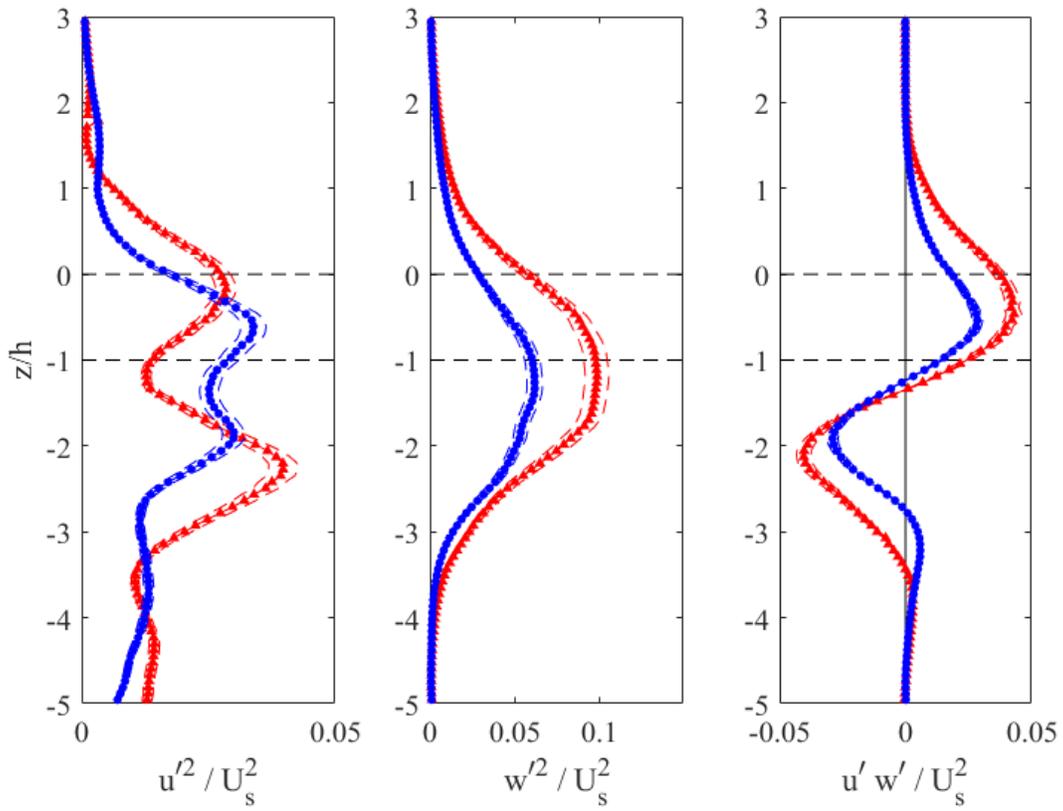

**Fig. 6** Fluctuating velocity profiles for the in-phase and out-of-phase models (see **Fig. 5** for legend). Shown are the streamwise ($u'^2$), vertical ($w'^2$) and cross-component ($u'w'$) profiles with dashed lines corresponding to uncertainty bands

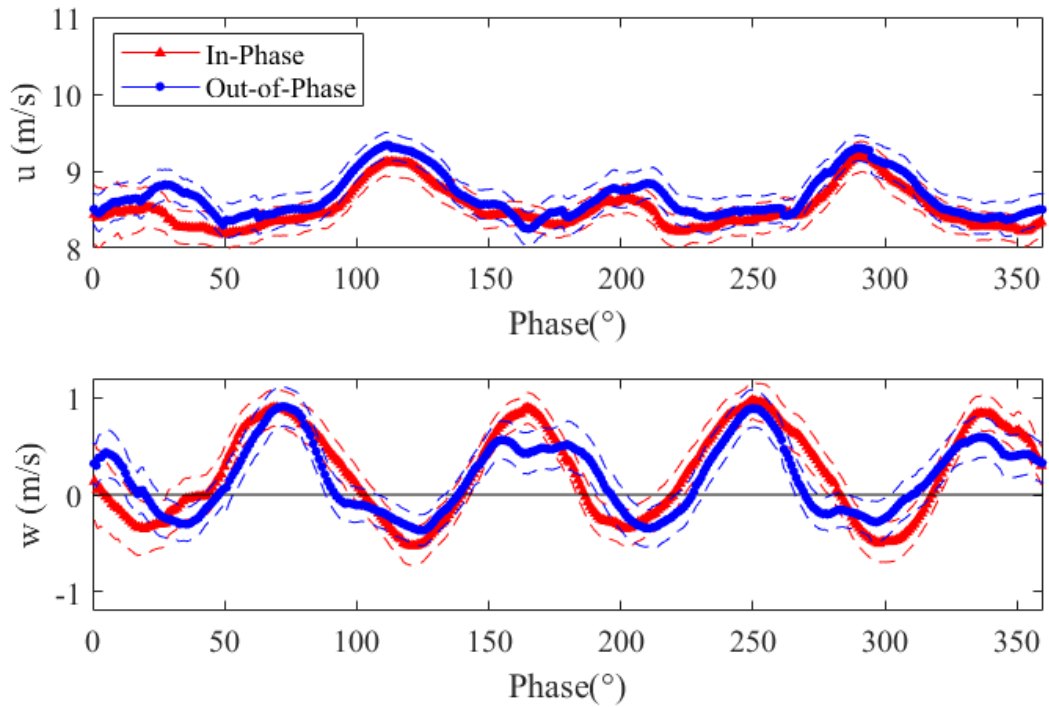

**Fig. 7** Example phase trace of the streamwise (*u*) and vertical (*w*) velocity from *z/h* = -2 for the in-phase and out-of-phase model. Dashed lines correspond to uncertainty determined for each measurement following the analysis of Wieneke (2015)

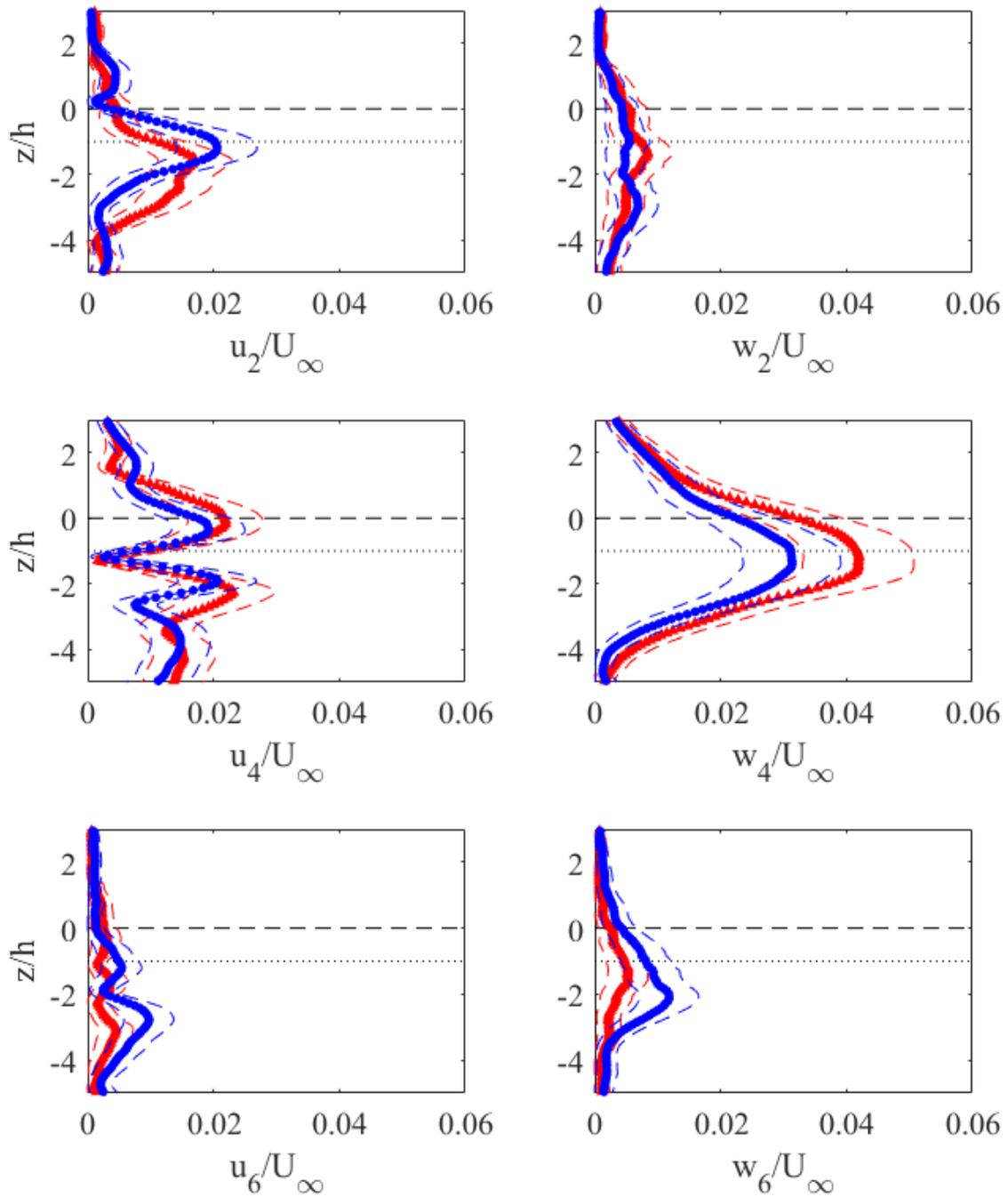

**Fig. 8** Model comparison of the vertical distribution of the (top row) 2/rev, (middle row) 4/rev and (bottom row) 6/rev spectral levels within the far wake from the (left column) streamwise and (right column) vertical velocity components. Dashed lines correspond to the estimated uncertainty bands

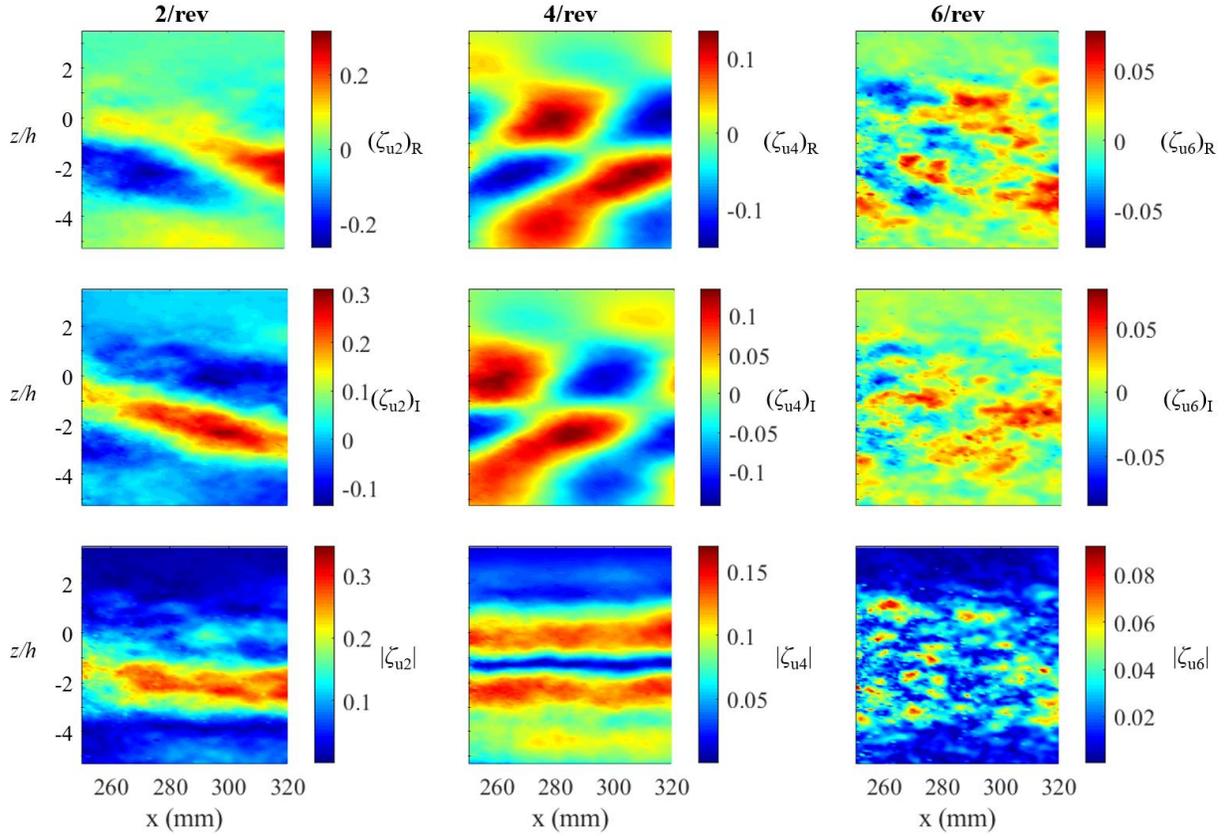

**Fig. 9** DMD analysis of the streamwise velocity within the wake of the in-phase model. Shown are the (top row,R) real, (middle row, I) imaginary and (bottom row) magnitudes of DMD modes ($\zeta$) corresponding to the (left column) 2/rev, (middle column) 4/rev and (right column) 6/rev frequencies

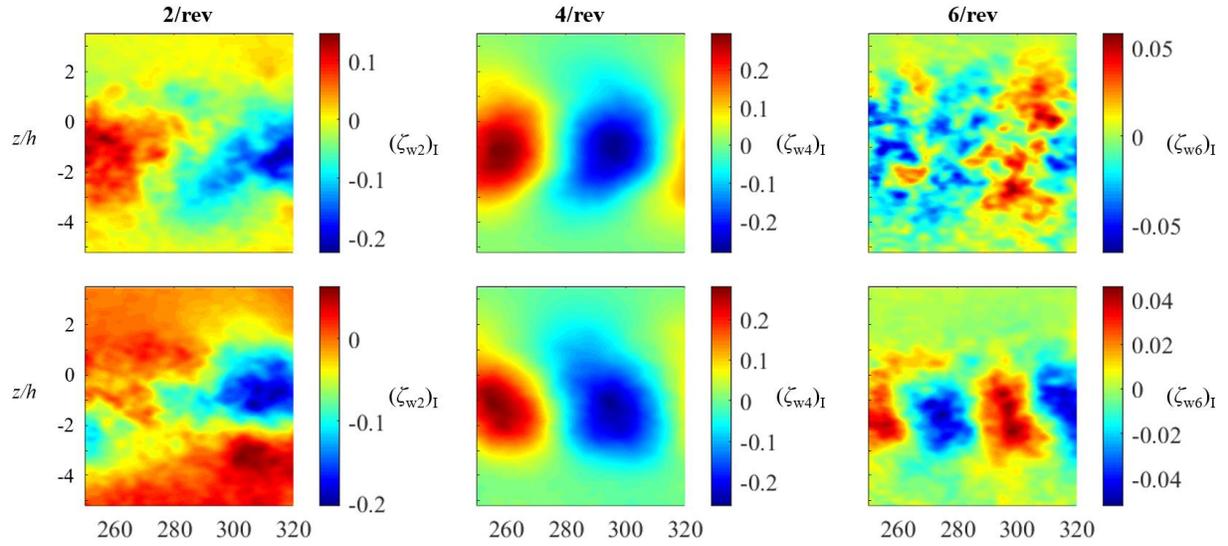

**Fig. 10** Comparison between the (top) in-phase and (bottom) out-of-phase models. Shown are the imaginary components of the DMD modes corresponding to the (left column) 2/rev, (middle column) 4/rev and (right column) 6/rev frequencies

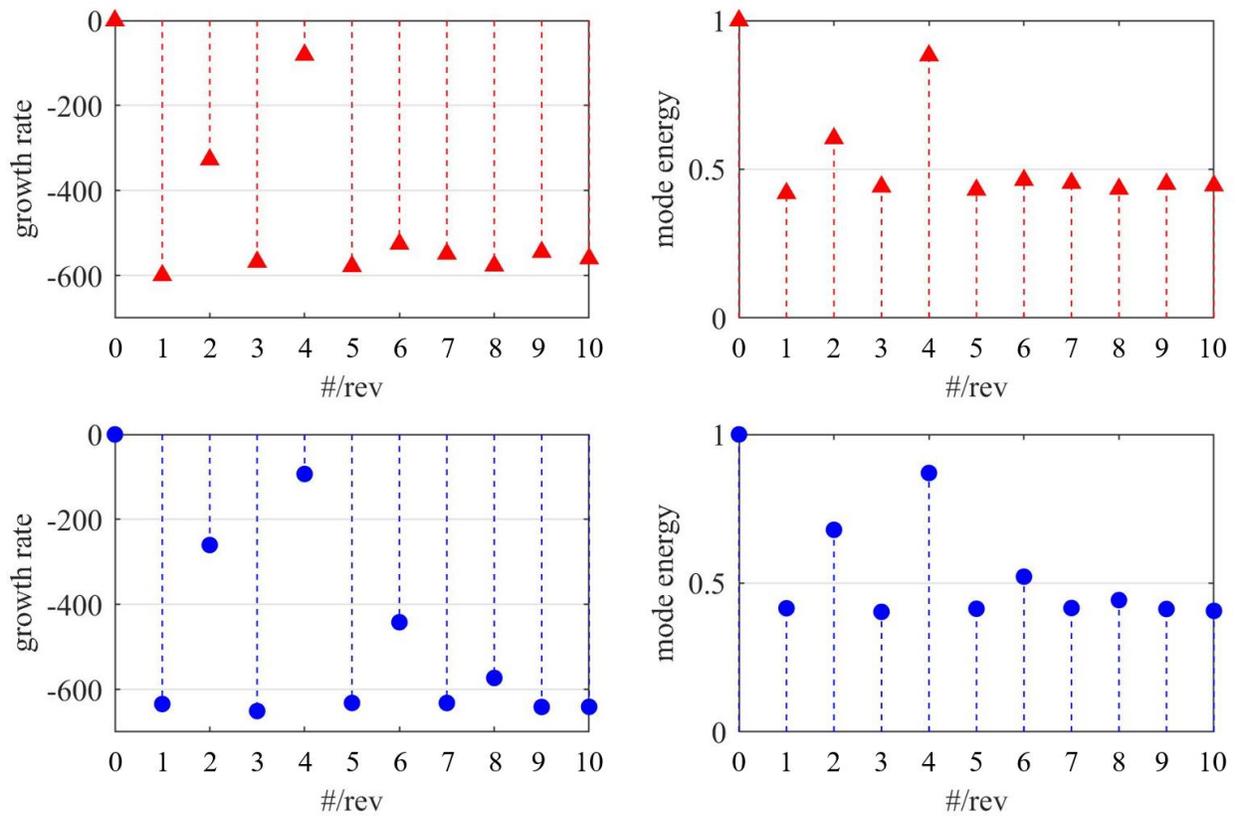

**Fig. 11** Comparison of the growth rate and mode energy for the (top) in-phase and (bottom) out-of-phase model

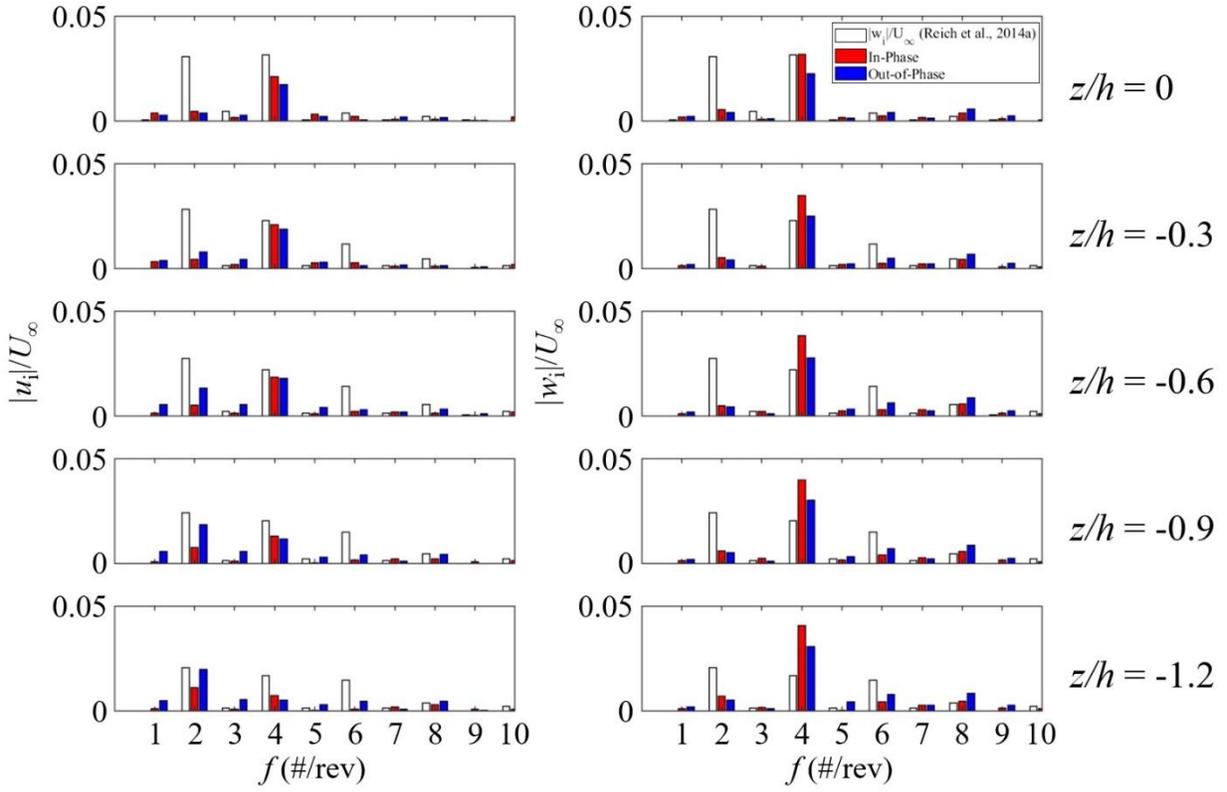

**Fig. 12** Spectral content from the (left) streamwise and (right) vertical velocity components scaled with the freestream speed from both models compared with the spectral levels from the vertical velocity component on a defeatured commercial model (Reich et al. (2014a)